# Structural Vibration Monitoring with Diffractive Optical Processors

Yuntian Wang[a,b,c], Zafer Yilmaz[d], Yuhang Li[a,b,c], Edward Liu[a], Eric Ahlberg[d], Farid Ghahari[e], Ertugrul Taciroglu[d], Aydogan Ozcan[a,b,c*]

[a]Electrical and Computer Engineering Department, University of California, Los Angeles, CA, 90095, USA

[b]Bioengineering Department, University of California, Los Angeles, CA, 90095, USA

[c]California NanoSystems Institute (CNSI), University of California, Los Angeles, CA, 90095, USA

[d]Civil and Environmental Engineering Department, University of California, Los Angeles, CA, 90095, USA

[e]California Geological Survey, California Department of Conservation, Sacramento, CA 95814, USA

[*]Correspondence to: ozcan@ucla.edu

## Abstract

Structural Health Monitoring (SHM) is vital for maintaining the safety and longevity of civil infrastructure, yet current solutions remain constrained by cost, power consumption, scalability, and the complexity of data processing. Here, we present a diffractive vibration monitoring system, integrating a jointly optimized diffractive layer with a shallow neural network-based backend to remotely extract 3D structural vibration spectra, offering a low-power, cost-effective and scalable solution. This architecture eliminates the need for dense sensor arrays or extensive data acquisition; instead, it uses a spatially-optimized passive diffractive layer that encodes 3D structural displacements into modulated light, captured by a minimal number of detectors and decoded in real-time by shallow and low-power neural networks to reconstruct the 3D displacement spectra of structures. The diffractive system's efficacy was demonstrated both numerically and experimentally using millimeter-wave illumination on a laboratory-scale building model with a programmable shake table. Our system achieves more than an order-of-magnitude improvement in accuracy over conventional optics or separately trained modules, establishing a foundation for high-throughput 3D monitoring of structures. Beyond SHM, the 3D vibration monitoring capabilities of this cost-effective and data-efficient framework establish a new computational sensing modality with potential applications in disaster resilience, aerospace diagnostics, and autonomous navigation—where energy efficiency, low latency, and high-throughput are critical.

## Introduction

Ensuring the safety and longevity of civil infrastructure, such as buildings, bridges, and dams, is paramount for societal well-being and economic stability[1–4]. Structural Health Monitoring (SHM) systems play a critical role in this endeavor by providing methods to assess the conditions of structures, detect damage, and predict remaining service life, particularly after exposure to natural hazards like earthquakes[5,6]. Traditional SHM systems often rely on visual inspections, which suffer from significant drawbacks: they require highly skilled personnel, can be time-consuming, costly, dangerous, and subjective, and may not always be feasible for inaccessible parts of a structure or large-scale assessments[7]. Alternative approaches include non-destructive testing (NDT)[8–11] and vibration-based SHM[12–15]. For example, vibration-based methods utilize sensor networks (e.g., accelerometers, strain gauges) to record structural responses to excitations. The analysis of this data, often through system identification techniques[16–37], aims to extract modal parameters (frequencies, damping ratios, mode shapes) whose changes can indicate damage. While powerful, conventional sensor networks can be expensive to install and maintain, require significant power, and generate large datasets demanding complex digital signal processing. Furthermore, achieving high spatial resolution for accurate damage localization often necessitates a dense and costly sensor deployment. Recent advancements like digital twins[38–43] leverage these methods but still depend on the quality and density of the underlying sensor data. Newer sensing technologies, including fiber optics[44], laser Doppler vibrometry[45,46], and vision-based systems[47,48], offer improvements but still face challenges in their relative costs, deployment complexity, or sensitivity.

Here, we present a structural vibration monitoring approach integrating diffractive processor-based encoders with shallow neural network-based decoders to accurately and rapidly reconstruct the 3D oscillation amplitudes and frequencies of a structure under test. In this architecture, an optimized encoder surface with wavelength scale diffractive features is attached to the structure of interest; as it oscillates at various frequencies, the corresponding movement of the passive diffractive layer modulates the reflected wavefront. This modulated light is captured by a few detectors, generating time-series signals that encode the 3D vibration spectra of the test structure. Shallow and low-power neural networks, co-optimized with the diffractive layer design, rapidly decode the detector signals to quantify the 3D displacement spectra of the structure. This diffractive 3D vibration monitoring approach offers the potential for a low-power, cost-effective, and scalable solution for parallel monitoring of structures. The passive nature of the structurally optimized diffractive layer reduces power requirements, while the computational capabilities of the deep learning-based backend enable robust decoding of vibration spectra from sparse detector data, also eliminating the acquisition and storage of large amounts of data. This co-optimization strategy ensures that the passive diffractive layer is specifically tailored to encode the structural displacements in a way that is optimally decodable by a shallow and low-power backend network, maximizing sensitivity and accuracy while also reducing complexity and digital data burden.

Our results and analysis demonstrate the validation of this diffractive vibration monitoring concept (Fig.1), revealing its accuracy in extracting the vibration spectra of different structures. We provide its experimental proof-of-concept using millimeter-wave illumination and a 3D-printed diffractive layer mounted on a laboratory-scale building model subjected to 1D and 2D dynamic excitations. We show that the jointly optimized diffractive system significantly outperforms other configurations using conventional optical elements, validating the efficacy of the jointly optimized design for high-fidelity vibration monitoring of structures with more than an order of magnitude improvement in accuracy.

The approach presented here marks a fundamental departure from conventional digital sensing paradigms by shifting a portion of the computational burden into the physical domain. This is achieved through the co-design and joint optimization of a passive diffractive encoder and a shallow, low-power neural network decoder. Unlike traditional sensor networks that digitize raw physical signals for subsequent extensive digital processing, our system leverages the diffractive layer as an optimized optical processor that intelligently pre-encodes complex, multi-dimensional structural oscillation information directly into modulated optical signals. This physical–digital co-integration provides a new foundation for real-time, energy-efficient computational sensing systems that are not limited to civil infrastructure. By reframing the problem of structural vibration monitoring as an optical inference task, our work introduces a compact, scalable, and cost-effective platform with broad relevance to distributed sensing in resource-limited settings, from smart cities and disaster prevention to aerospace diagnostics and autonomous navigation. These results represent not just a new tool for SHM, but a novel modality in physical computation and signal encoding, capable of transforming how we remotely sense and interpret dynamic physical systems. Furthermore, this system exemplifies the principles of edge AI in the physical domain, where hardware is not merely a conduit for digital inference, but an active and co-optimized participant in the computational process itself. Such advances could lay the groundwork for future systems that seamlessly integrate sensing, encoding, and inference into unified, compact optical layers—paving the way for scalable networks of intelligent sensors deployed at an unprecedented scale across various scientific and engineering disciplines.

## Results

**Diffractive systems for monitoring 3D structural vibrations**

We designed a diffractive system capable of measuring and extracting 3D structural oscillations using an optimized reflective layer. The system configuration is depicted in Fig. 2a. Input wave, incident at an oblique angle relative to the normal, propagates through free space onto the reflective diffractive layer (shown in Fig. 2b). This reflective diffractive surface, affixed to the target structure under test, consists of a 200×200 array of trainable, phase-only diffractive features, each approximately ~$\lambda/2$ in lateral size, where $\lambda$ is the illumination wavelength. Mechanically attached to the structure, this passive diffractive layer follows the structure's displacements, modeled as a linear combination of various harmonics in x, y and z directions with randomly generated amplitudes and phases (see Methods for details). The reflected wave that is spatio-temporally modulated by the optimized diffractive layer propagates to the output plane and is sampled by four detectors at 50 Hz. The detector array contains four single-pixel detectors arranged uniformly, with a center-to-center spacing of 16$\lambda$ (Fig. 2c). The detected time signals serve as an encoded input to a shallow displacement decoder network that estimates the 3D structural displacement time series, i.e., x($t$), y($t$), and z($t$). A subsequent frequency processor network then analyzes the displacement data to extract the 3D oscillation spectra, covering a pre-determined range of 9 to 11 Hz (selected, without loss of generality, as the target band of interest) in x, y and z directions (Fig. 2a).

    The diffractive layer's phase profile with a lateral pitch of ~$\lambda/2$ and the digital parameters of the backend neural networks were co-optimized through a joint training procedure (detailed in Methods) to maximize the accuracy of the 3D spectral reconstructions. For quantitative comparison, we also used additional optical

elements as baseline designs to compare with the jointly optimized diffractive system. Specifically, we evaluated a separately optimized diffractive layer, a Fresnel lens array and a random phase diffuser (see Methods for details). In this comparison, the separately trained diffractive layer had the same number of optimizable diffractive features, and its surface profile was optimized to efficiently communicate with and focus light onto the output detector array; see Fig. 3a. Importantly, the backend neural network architecture, including the displacement decoder network and the frequency processor network, remained the same across all the configurations used in this comparison; in these cases, the backend networks were trained separately for accurate reconstruction of the vibration spectra of the structure after the corresponding optical components were fixed.

The performance comparison of these different designs is shown in Fig. 3b, where the input spectra and the reconstructed spectra for each configuration are compared to each other. Performance was quantified using the Mean Squared Error (MSE) of the spectrum within the target frequency range, i.e., 9 Hz to 11 Hz. The performance of the 3D spectral inference for these four configurations is also reported in **Table 1** as a function of the number of trainable parameters ($N_D$) within the displacement decoder network. The jointly optimized diffractive layer consistently achieved the lowest MSE, with more than an order of magnitude improvement over the other configurations across different model capacities. While some of the alternative configurations showed some predictive capability, they exhibit significantly higher spectral reconstruction errors, as reported in Fig. 3 and Table 1, even with a larger number of trainable parameters in the digital backend. On the other hand, the diffractive vibration monitoring system, with the jointly optimized optical and digital processors, revealed a significantly better 3D spectral inference performance.

The superior performance of the jointly trained SHM system is further demonstrated through its blind testing performance in single-frequency extraction (in 3D), as detailed in the confusion matrices reported in

Fig. 4a. In this analysis, various single-frequency oscillations were applied to the structure under test along one of the directions within the target spectral range. The diffractive system's inference results, denoted as $\hat{x}, \hat{y}$ and $\hat{z}$, corresponding to these single frequency inputs are shown as columns in the confusion matrix (Fig. 4a). The presence of a clear diagonal in all the configurations of the jointly optimized diffractive layer represents the accurate prediction of these single harmonic oscillations in all 3 directions (x, y, z) with minimal crosstalk between neighboring frequencies; on the other hand, significant spectral crosstalk or even complete failures (especially for oscillations in depth, i.e., the z direction) are observed in the alternative configurations as shown in the corresponding confusion matrices reported in Fig. 4a. Apart from these 3D analyses, we also obtained similar results for 2D vibration analysis using diffractive processors as reported in Supplementary Figs. S1-S2 and Supplementary Table S1.

| **Spectral MSE** | $N_D = 2.98k$ | $N_D = 1.75k$ | $N_D = 0.85k$ |
|---|---|---|---|
| Diffractive layer (jointly optimized) | $1.109 \times 10^{-2}$ | $1.400 \times 10^{-2}$ | $1.654 \times 10^{-2}$ |
| Diffractive layer (separately optimized) | $1.419 \times 10^{-1}$ | $1.675 \times 10^{-1}$ | $2.016 \times 10^{-1}$ |
| Fresnel lens array | $3.576 \times 10^{-1}$ | $3.834 \times 10^{-1}$ | $4.264 \times 10^{-1}$ |
| Random diffuser | $6.243 \times 10^{-1}$ | $6.545 \times 10^{-1}$ | $6.760 \times 10^{-1}$ |

**Table 1** 3D spectral MSE results of different optical configurations, evaluated as a function of the number of trainable parameters ($N_D$) of the displacement decoder network.

While the joint optimization of the diffractive encoder and the digital backend network demonstrated significant performance advantages, as reported above, a common objective is to develop more compact models that require less computational resources, thereby enhancing potential deployability. This practical consideration motivates a deeper investigation into the relationship between the inference model complexity and performance. To better understand this relationship, we analyzed the trade-off between the displacement decoding network complexity ($N_D$) and 3D spectral reconstruction performance. For this analysis, we varied the network's hidden layer dimensions and examined the correlation between the total floating-point operations (FLOPs) and the resulting 3D spectral MSE. As shown in Fig. 4b, an increased model capacity with a larger $N_D$ generally leads to improved spectral inference accuracy, quantifying the performance trade-off between the spectral MSE and the computational cost; these observations are also in agreement with our results reported in Table 1.

**Experimental demonstration of diffractive vibration monitoring**

We experimentally validated the presented concept using 3D-printed diffractive layers and millimeter wave illumination. An initial experimental setup, depicted in Fig. 5a, was designed to monitor the 1D (x-axis) oscillations of a test structure, which was a four-level building model with its base floor mounted on a programmable shake table for controlled perturbations. This experimental validation is not only a performance check but also a demonstration of how physical encoding combined with minimal-data processing can operate in noisy, non-ideal conditions—a key benchmark for deployment in uncontrolled environments. To suppress displacements in the z-direction and concentrate motion primarily along the x-axis, metal wires were used to constrain the model along the z-axis, effectively increasing its stiffness in that direction. Ground truth displacement data for the test structure were simultaneously acquired using laser rangefinders. The core of the diffractive vibration monitoring system involved an optimized, 3D-printed diffractive layer, which was affixed to the first level of the building model (Fig. 5a). A coherent source at $\lambda = 3\ mm$ illuminated the structure, and the reflected waves were spatio-temporally modulated by the displacements of the passive diffractive layer on the structure; these reflected signals were captured by two (single-pixel) detectors (see Methods section). During these measurements, various excitation methods were used to induce different structural vibrations (shown in Fig. 5b), including applying a white noise signal to the shake table, programmed shaking profiles of the base floor, and manual perturbations to the base, first and upper levels. Furthermore, to simulate variations in the building structure, a mass block was systematically placed and fixed at different levels, altering both the mass distribution and the dynamic response of the system, thereby changing its natural frequency.

The temporal signals captured by the two detectors were decoded by a trained digital backend, which predicted the frequency spectra of the structure's vibrations within a predefined spectral range of interest. Figure 6 and Supplementary Fig. S3 compare the ground truth spectra (measured by a laser rangefinder) and the spectra extracted by the diffractive system for various structures with different mass placements and perturbations. The accurate predictions of both the vibration frequencies and their corresponding amplitudes in these experimental results confirm the feasibility and effectiveness of the diffractive system for monitoring structural dynamics. We also compared the inference performance of the optimized diffractive layer to that of a reflective flat mirror, as depicted in Supplementary Fig. S4. In this comparison, both sets of the digital backend neural networks were trained/optimized using an identical number of measurements with the same network architecture (i.e., with the same number of trainable parameters) to provide a fair comparison. The configuration with the optimized diffractive layer yielded a significant improvement in its spectral MSE results over the results achieved with a flat mirror, which stems from the optimized diffractive layer's ability to interact more effectively with the output detectors. Compared to a flat mirror or another reflective optical element, an optimized diffractive layer enhances the standard deviation of the detected signals under various structural vibrations. This results in an enhanced 3D oscillation encoding capability through the diffractive layer since a higher standard deviation in the detected signals means that different oscillation spectra produce more distinguishable intensity distributions across the output detectors, making it easier for the jointly optimized digital decoder to learn the mapping required for accurate spectral reconstructions.

To further improve the spectral inference accuracy of our approach, we developed a post-processing method based on temporal averaging, outlined in Fig. 7a. This strategy employs a sliding window of constant duration across each detector's temporal signal stream to generate multiple (potentially overlapping) time-series segments. The range of the sliding windows is defined by the temporal averaging time $\Delta_t$, representing the total temporal span across which the centers of the sliding windows are distributed for the averaging calculation. Each segment serves as an individual input to the trained digital backend, which extracts the oscillation spectrum associated with that time window. The final spectral output is then computed by averaging the ensemble of vibration spectra extracted from all of the time segments. The inherent trade-off between spectral MSE and the cumulative FLOPs is depicted in Fig. 7b; averaging over a greater number of temporal windows improved the spectral inference accuracy (lowering the spectral MSE) while also increasing the required computational time. Figure 7c provides a comparison of the extracted spectra for different $\Delta_t$ values, illustrating the impact of this approach on the quality of our spectral reconstructions.

We further investigated the diffractive processor-based vibration monitoring system by measuring 2D (x and z) oscillations of the same building model (see Fig. 8a). For these experiments, the restraining wires that were previously used to prevent motion along the z-axis were loosened to allow for 2D displacements of the test structure. Laser rangefinders were installed on adjacent sides of the first level (shown in Fig. 8b) to provide reference (ground truth) measurements of 2D displacements. Another optimized diffractive layer (shown in Fig. 8a), mounted on the same level, was used to spatio-temporally modulate the incident wave in accordance with the structure's 2D vibrations. The test structure was subjected to 2D oscillations using a programmable shake table that reproduced seismic waveforms in the x-direction from the NGA-West2 earthquake dataset[49], applied at both full scale and 0.1x scale while the perturbations of the z-axis were introduced by pushing manually (details in Methods). The diffractive processor-based system simultaneously extracted the spectra of the building's displacements in both x and z directions from the 2 detector signals. Examples of spectral inference results for x- and z-directions are visualized in Fig. 8c, along with the corresponding ground truth measurements. The performance advantages of the optimized diffractive layer compared to the 2D inference results obtained with a flat mirror are also illustrated in Fig. 8d, supporting the same conclusions as in our earlier experiments reported in Supplementary Fig. S4.

The performance advantages observed through the joint optimization of the diffractive layer and the digital backend extends beyond mere accuracy improvements; it represents a new approach to how sensing systems can be designed. By co-designing the passive optical encoder material with the subsequent shallow neural network decoder, the system implicitly learns to transform complex, multi-dimensional structural oscillation information into highly distilled, yet robustly decodable, optical signals that can be captured by a minimal number of detectors. This optical pre-processing effectively offloads significant computational burden from the digital domain to the physical layer, enabling the use of shallow and low-power neural networks for rapid and accurate spectral reconstruction, a capability unattainable with traditional sensing architectures that would require far more detectors and extensive post-processing for comparable performance. This fundamental principle of learned physical-digital compression holds transformative potential for numerous sensing applications where data efficiency, low power, and remote operation are critical constraints, such as in autonomous navigation and distributed environmental sensing. The synergy between the optimized physical-layer encoding and the efficient neural network decoding signifies a move towards "intelligent sensing at the source," where essential information is extracted and compacted through light-matter interactions before traditional digital processing takes over. This minimizes data and computational overhead, addressing a critical bottleneck in ubiquitous, high-density sensor deployments and laying the groundwork for truly distributed, low-latency monitoring networks.

## Discussion

We developed a hybrid system, integrating a jointly optimized diffractive layer with a shallow digital neural network backend for rapidly extracting structural vibration frequency spectra. The effectiveness of this hybrid system was demonstrated through both numerical simulations and experiments using millimeter-wave illumination. We observed major performance improvements through the joint optimization of the diffractive layer and the backend neural networks; this jointly-optimized system not only significantly outperformed other optical configurations with non-trainable optical elements such as flat mirrors, Fresnel lens arrays, or random diffusers, but also showed major performance improvements compared to separately optimized diffractive layers. These observations can be attributed to the jointly optimized diffractive layer's ability to effectively encode 3D structural vibrations into the optical wavefront, enhancing signal variations at the output detectors and creating more distinguishable spatio-temporal patterns in response to various structural perturbations, helping the decoder backend to reveal the 3D oscillation spectra accurately.

The significance of this work extends beyond potential performance gains in SHM. The use of trainable diffractive optics as a front-end signal encoder, paired with a jointly trained neural decoder, introduces a new strategy for compact, low-data-rate sensors that can operate without dense sampling or extensive storage. This hybrid design paradigm may inspire new classes of passive sensors in disciplines such as aerospace engineering and robotics, where power and size constraints preclude conventional sensing architectures.

Beyond demonstrating the core functionality of diffractive vibration monitoring systems, this work also investigated practical trade-offs influencing the system's performance and computational demands. We quantitatively explored the relationship between the complexity of the digital backend network and spectral reconstruction accuracy by varying the hidden layer dimensions of the displacement decoding network. As illustrated in Fig. 4b, employing decoder networks with more trainable parameters, corresponding to higher numbers of FLOPs, generally leads to reduced 3D spectral MSE. Additionally, we introduced and experimentally validated a temporal averaging-based post-processing strategy, which leverages multiple time segments from the sensor stream to enhance the final spectral reconstruction fidelity significantly. These findings provide valuable insights for tailoring the diffractive system's configuration to specific application requirements, balancing accuracy, latency, and available computational power. A key aspect of the temporal averaging-based post-processing method is its departure from traditional linear operations, such as those employed in standard periodograms. Our technique utilizes a non-linear neural network-based displacement decoder, which interacts differently with the diverse features present in individual segments of the time-sequence detector data. This processing through the network's architecture underpins a complex N-to-N non-linear mapping from the raw temporal segments to their respective spectral estimates. Such temporally shifted successive non-linear mappings offer distinct advantages, particularly in the capacity to more effectively mitigate noise and handle unexpected or out-of-distribution data points that may be present in the measured time signals.

While our experimental validation at millimeter-wave frequencies on a laboratory-scale model demonstrates the foundational principles and remarkable efficacy of this diffractive vibration monitoring system, the path toward widespread real-world deployment also presents unique challenges and opportunities that warrant careful consideration. Translating the millimeter-wave results to optical or infrared wavelengths for civil infrastructure applications will necessitate the development of large-area, robust diffractive layers with feature sizes on the order of hundreds of nanometers, resilient to environmental factors such as temperature fluctuations, humidity, and structural deformation over long periods. Furthermore, ensuring the long-term stability and calibration of both the diffractive layer and the remote illumination/detection system in dynamic outdoor environments will be crucial for maintaining the high accuracy demonstrated in controlled settings. The development of self-calibrating mechanisms or adaptive learning algorithms to compensate for environmental drift and material degradation will be important steps toward robust field implementations.

A significant future direction in diffractive SHM systems involves scaling this technology for higher throughput assessment of structures by monitoring numerous locations simultaneously. Extending the current approach for this goal could involve spatial multiplexing, where multiple co-designed monochrome diffractive vibration processors monitor different points, potentially feeding data into a unified backend neural network. An even more advanced concept involves spectral-spatial multiplexing, assigning distinct wavelengths to different monitored locations of a structure, each potentially utilizing a specifically tuned diffractive layer and detector array, allowing for potentially higher density of structures to be monitored in parallel and processed by a common backend decoder.

The exploration of shorter operational wavelengths, particularly within the visible and IR spectra, presents a compelling avenue for enhancing the accuracy and resolution of diffractive SHM systems due to the scalability of diffractive optical processors with respect to the wavelength of illumination. While the presented work experimentally validated the diffractive SHM systems using millimeter waves, the underlying principles of diffractive layer design allow for scalability across different parts of the electromagnetic spectrum by adjusting the dimensions of the diffractive features proportional to the illumination wavelength. Transitioning the illumination wavelengths to the visible or infrared spectrum would require the fabrication of diffractive elements with significantly smaller lateral feature sizes. This refinement in feature resolution, achievable with advanced 3D manufacturing techniques such as two-photon polymerization and optical lithography, offers the potential for more precise wavefront modulation[50–52]. The capability to produce such fine structures is critical for operating effectively at shorter wavelengths and could lead to substantial improvements in the sensitivity and overall accuracy of diffractive SHM systems. These advances could pave the way for a new generation of high-accuracy, compact, multiplexed and more cost-effective SHM solutions.

Given its foundation in diffractive optics and neural computation, our system is inherently adaptable to other spectral bands, functional targets, and time-varying environments. The potential to extend this approach into the visible or near-infrared regime—enabled by nano-scale fabrication—could enable ultra-compact sensors for environmental monitoring or industrial robotics. We envision that the physical encoding strategies described here will find broader applications across science and engineering, enabling systems where sensing and computation are not considered separately, but co-optimized within the same physical substrate.

## Methods

**Training Data Preparation**

The displacements along the axes were synthesized by a linear combination of harmonic oscillations in the target frequency window. The amplitudes of the harmonic oscillations were randomly generated with an amplitude upper bound of $A_{max}$:

$$A_{\{x,y,z\},j} \sim U[0, A_{max}]$$

where $A_{max}$ was set as $\sim 1.5\lambda$ while the phase values of the harmonic oscillations $\phi_{\{x,y,z\},j}$ were randomly generated between $0$ and $2\pi$. The displacement of the diffractive layer at time stamp $t$ was calculated using:

$$\Delta\{x,y,z\}_t = \sum_{j=1}^{M} A_{\{x,y,z\},j} \sin(2\pi f_j t + \phi_{\{x,y,z\},j})$$

where $M$ represents the total number of discrete frequencies considered. $f_j$ denotes the j-th frequency component, selected from the discrete set {8.0, 8.2, ..., 11.8, 12.0} Hz.

## Supplementary information

This file contains:

- Supplementary Table S1
- Supplementary Figures S1-S4.
- Numerical Forward Model
- Network Structure for the Digital Backend
- Training Scheme and Loss Function
- Experimental Design and Testing

# Figures

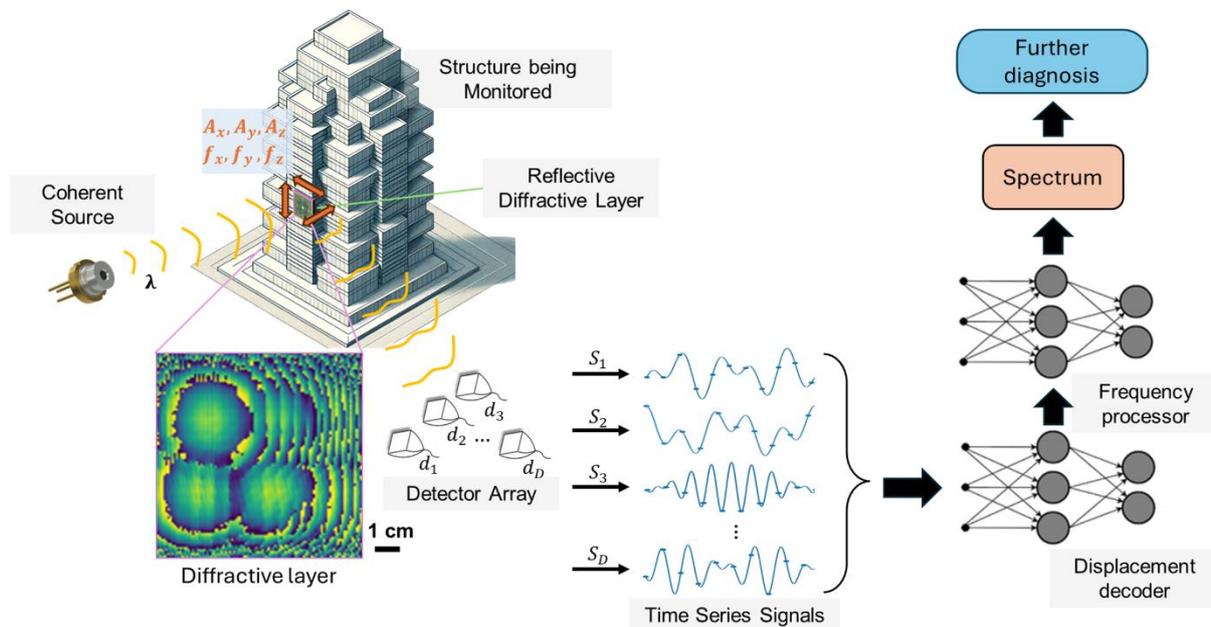

**Figure 1. The schematic of a diffractive optical processor designed for high-throughput structural vibration monitoring with a shallow electronic neural network-based decoding.** The diffractive optical processor, optimized using deep learning, is attached at various locations on a target structure, where it encodes local oscillatory motion into wavelength-dependent structurally optimized diffraction patterns. A few detectors capture the reflected light, which varies in signal strength based on the displacement amplitudes and frequencies at each point that is being monitored. The resulting time series signals are rapidly decoded by a shallow neural network backend (including a displacement decoder and a frequency processor) to simultaneously extract 3D oscillation features of the structure under test. This integrated system can enable cost-effective, low-power structural health diagnostics powered by deep learning-designed diffractive processors.

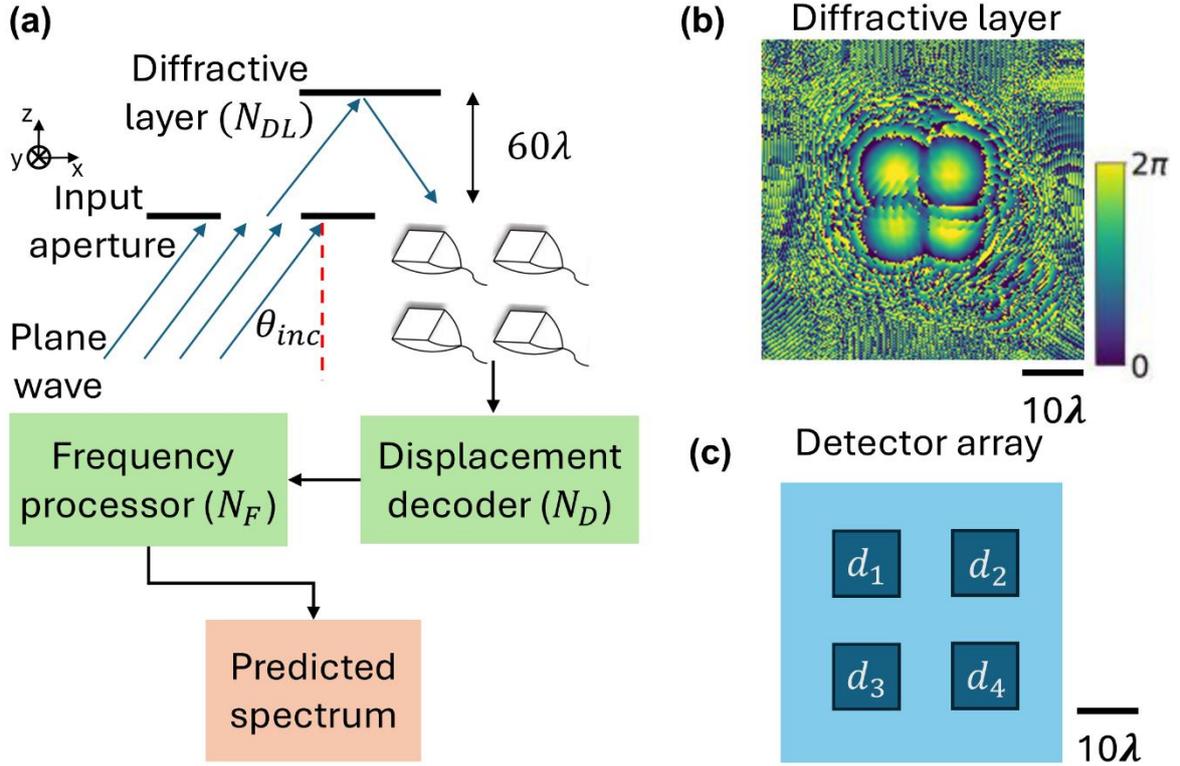

**Figure 2. Configuration and components of a diffractive vibration monitoring system.** (a) System configuration: an input wave (incident at $\theta_{inc}$) reflects off of an optimized diffractive layer, which is attached to an oscillating structure. The resulting optical signal (in reflection) is captured by a detector array and rapidly processed by a displacement decoder and a frequency processor, which measure the displacement and the 3D oscillation spectrum of the structure under test. $N_{DL}$, $N_D$ and $N_F$ are the number of trainable parameters in the optimized diffractive layer, displacement decoder network and frequency processor network, respectively. (b) The optimized phase modulation pattern of the trained surface of the diffractive vibration monitoring system. The diffractive processor phase profile was jointly trained with a displacement decoder that has $N_D = 6.39k$ trainable parameters. (c) Detector array layout with four single-pixel detectors.

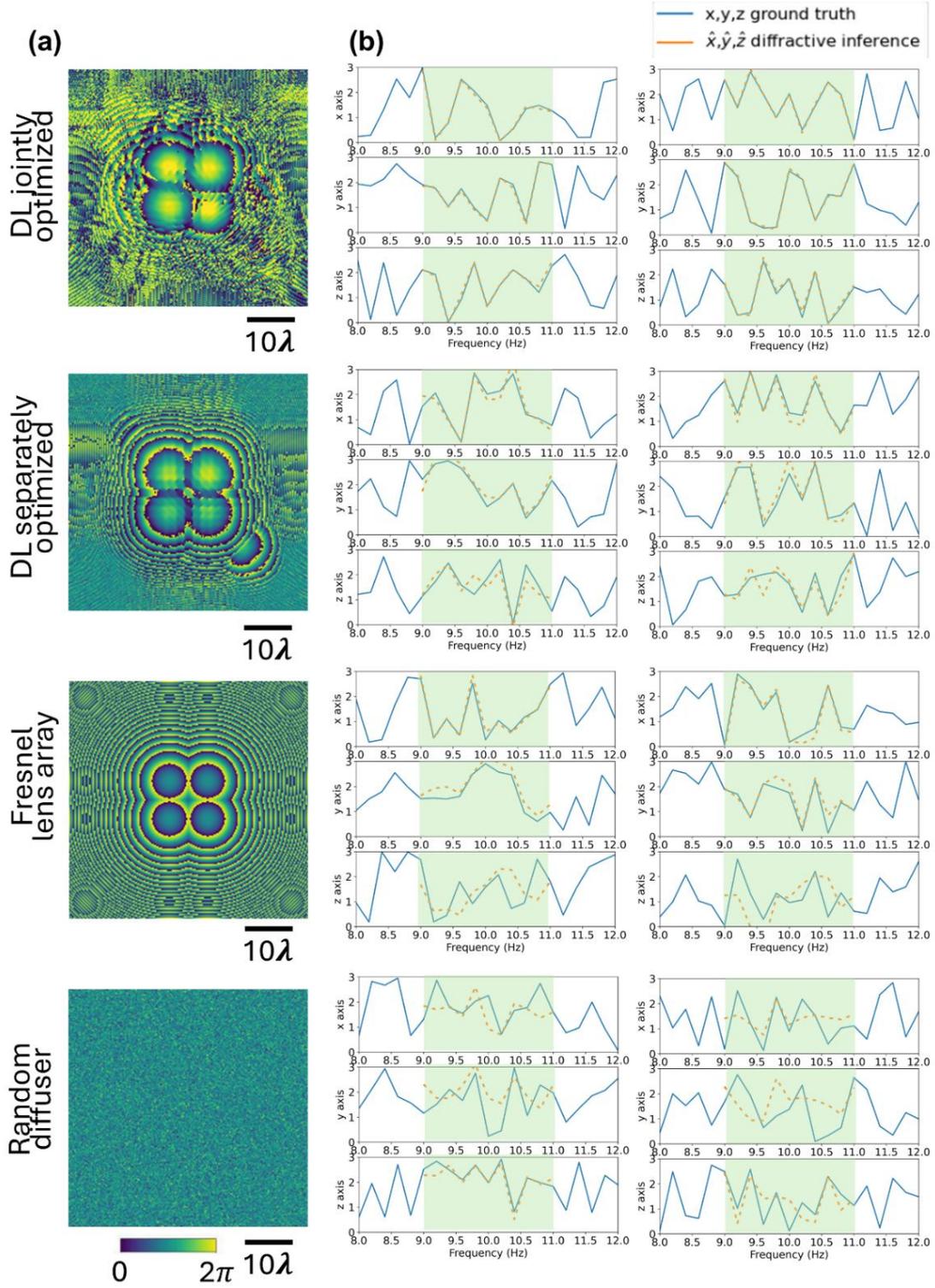

**Figure 3. Comparison of 3D oscillation spectra inference performance across different optical configurations; $N_D = 2.98k$.** (a) Phase modulation patterns of a jointly trained diffractive layer, a separately trained diffractive layer, a Fresnel lens array and a random phase diffuser, displayed from top to bottom, respectively. (b) Ground truth 3D oscillation spectra (two examples in each direction) and the diffractive inference results for each configuration. The green-shaded region highlights the frequency band of interest (9-11 Hz), i.e., the training range of the spatial oscillations. The spectral MSE values corresponding to the phase modulation patterns reported in the first column of Table 1 are $1.109 \times 10^{-2}$

(jointly optimized DL), $1.419 \times 10^{-1}$ (separately optimized DL), $3.576 \times 10^{-1}$ (Fresnel lens array), $6.243 \times 10^{-1}$ (random diffuser). The jointly optimized diffractive layer consistently achieved the lowest spectral MSE, with more than an order of magnitude improvement over other configurations – all of which used $N_D = 2.98k$ at the digital backend.

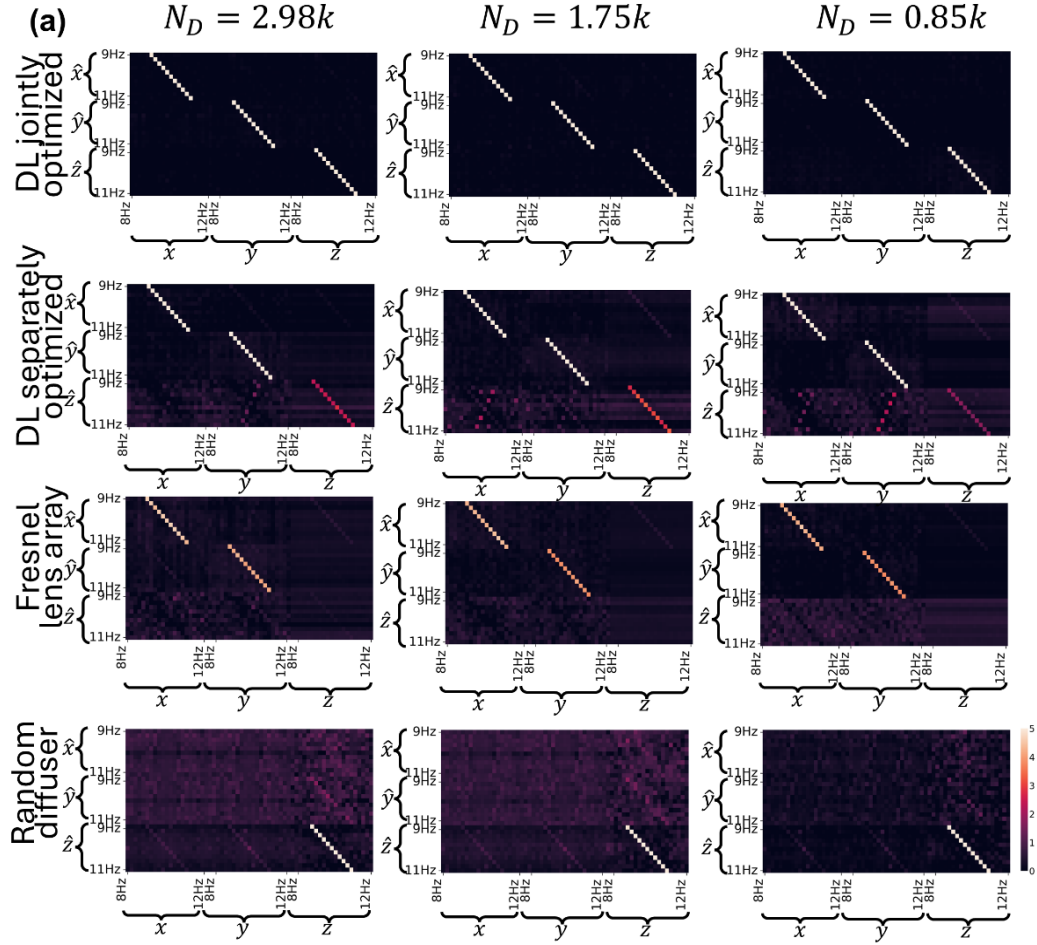

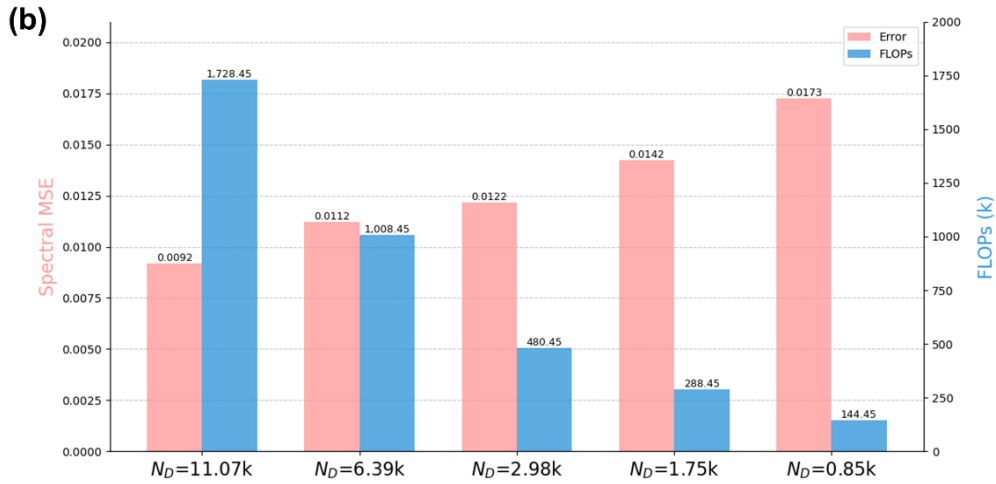

**Figure 4. Analysis of 3D oscillation spectra inference performance.** (a) Confusion matrices for single frequency inference across various system configurations and displacement decoder sizes ($N_D$). The horizontal axis represents input frequencies (ground truth), and the vertical axis represents the inference spectra (9-11Hz) for structural oscillations in x, y, and z. The color bar shows the inference intensity. (b) The trade-off between the decoder network complexity and the spectral MSE. Increased model capacity with a larger $N_D$ leads to improved spectral accuracy at the cost of an increase in the number of FLOPs needed.

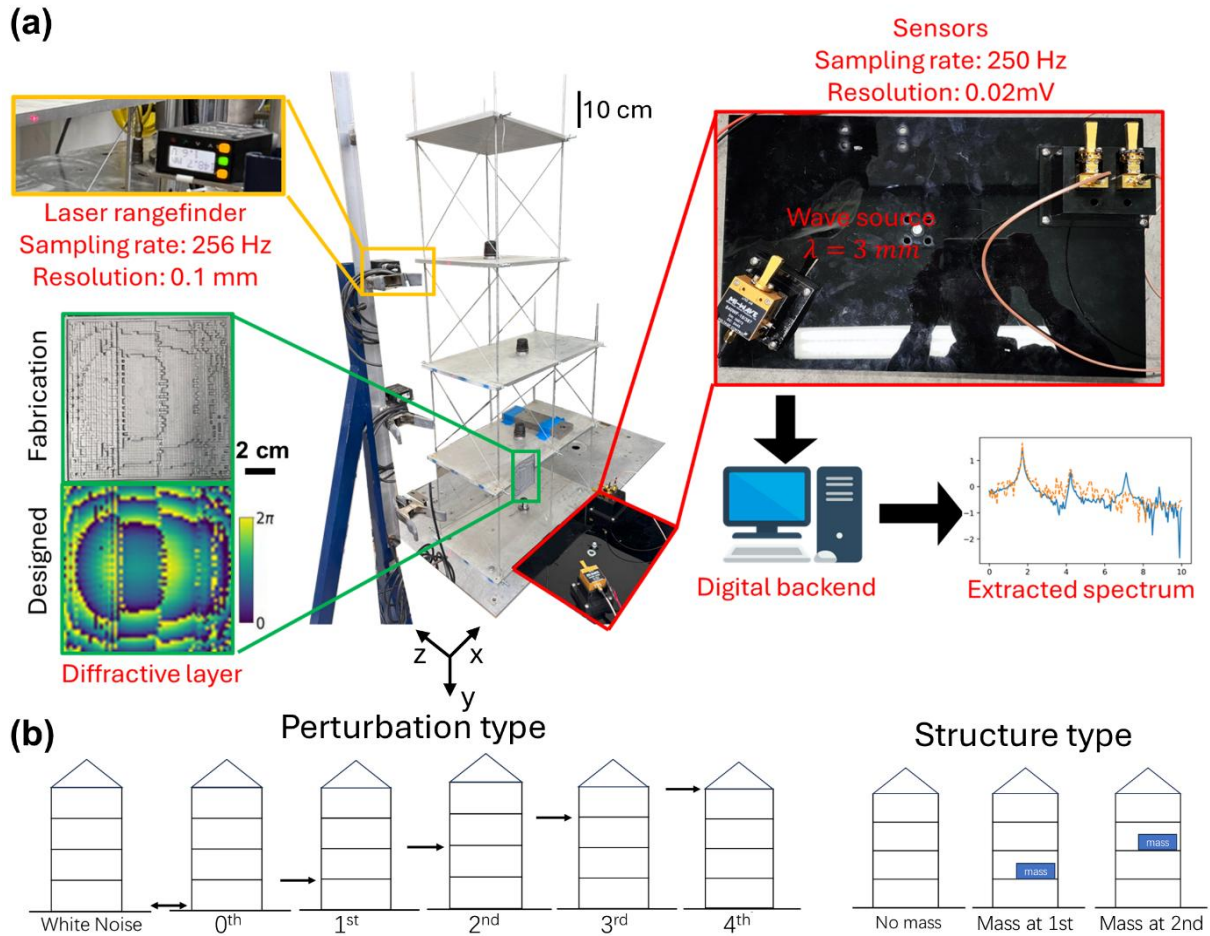

**Figure 5. Experimental validation of the 1D diffractive vibration monitoring system using a millimeter-wave ($\lambda = 3\ mm$) source.** (a) Photograph of the experimental setup and the pipeline for 1D spectrum inference. (b) Schematic of the types of perturbations and structures tested in these experiments.

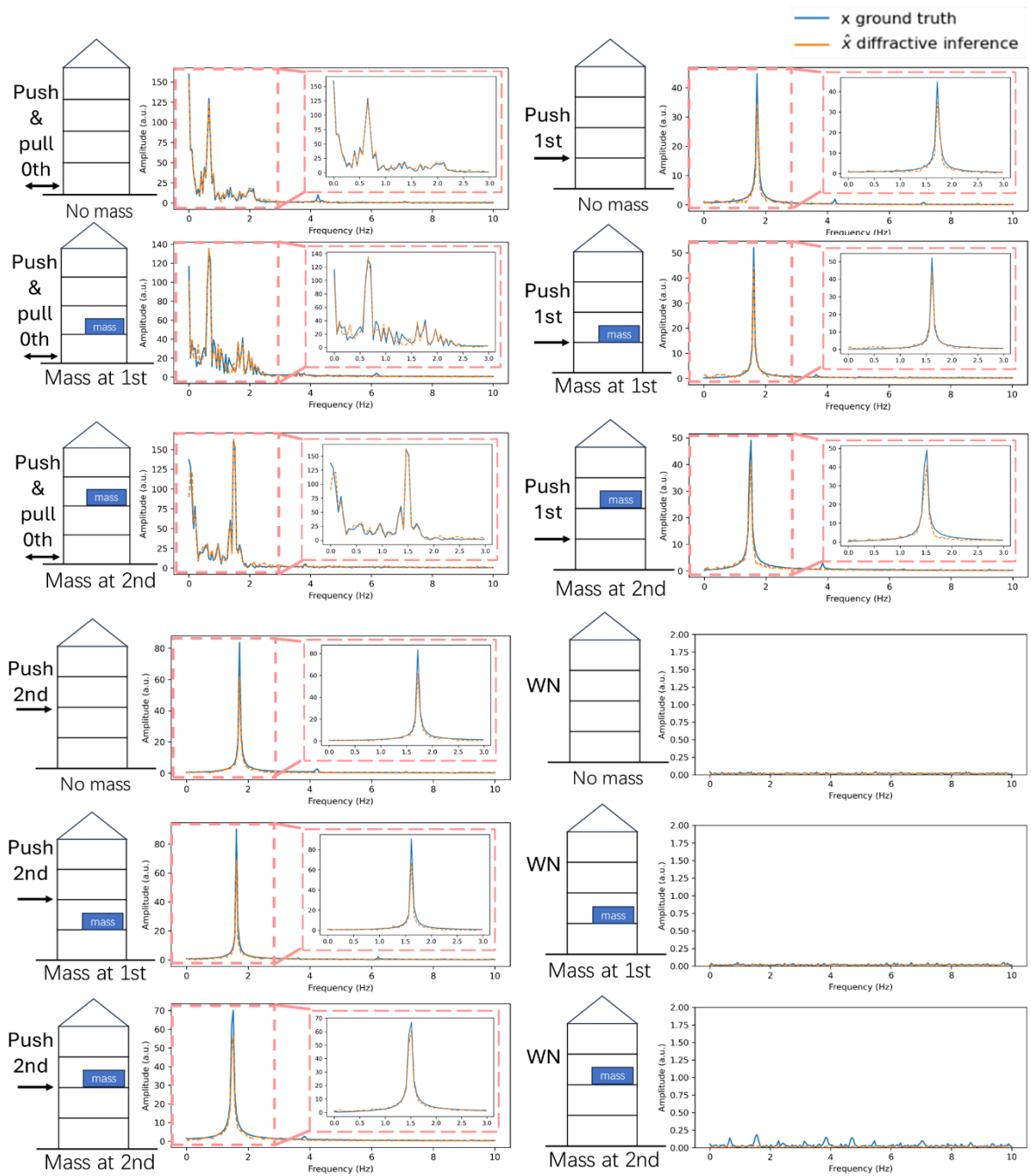

**Figure 6. Experimental results of the 1D diffractive vibration monitoring system using a millimeter-wave source.** Spectral inference results of different configurations with various types of perturbations and structures are compared against the ground truth. WN: white noise. The non-zero energy at zero frequency was caused by the displacement of the base level (0[th] level) due to the manual perturbation. Also see Supplementary Figs. S3-S4 for additional experimental results.

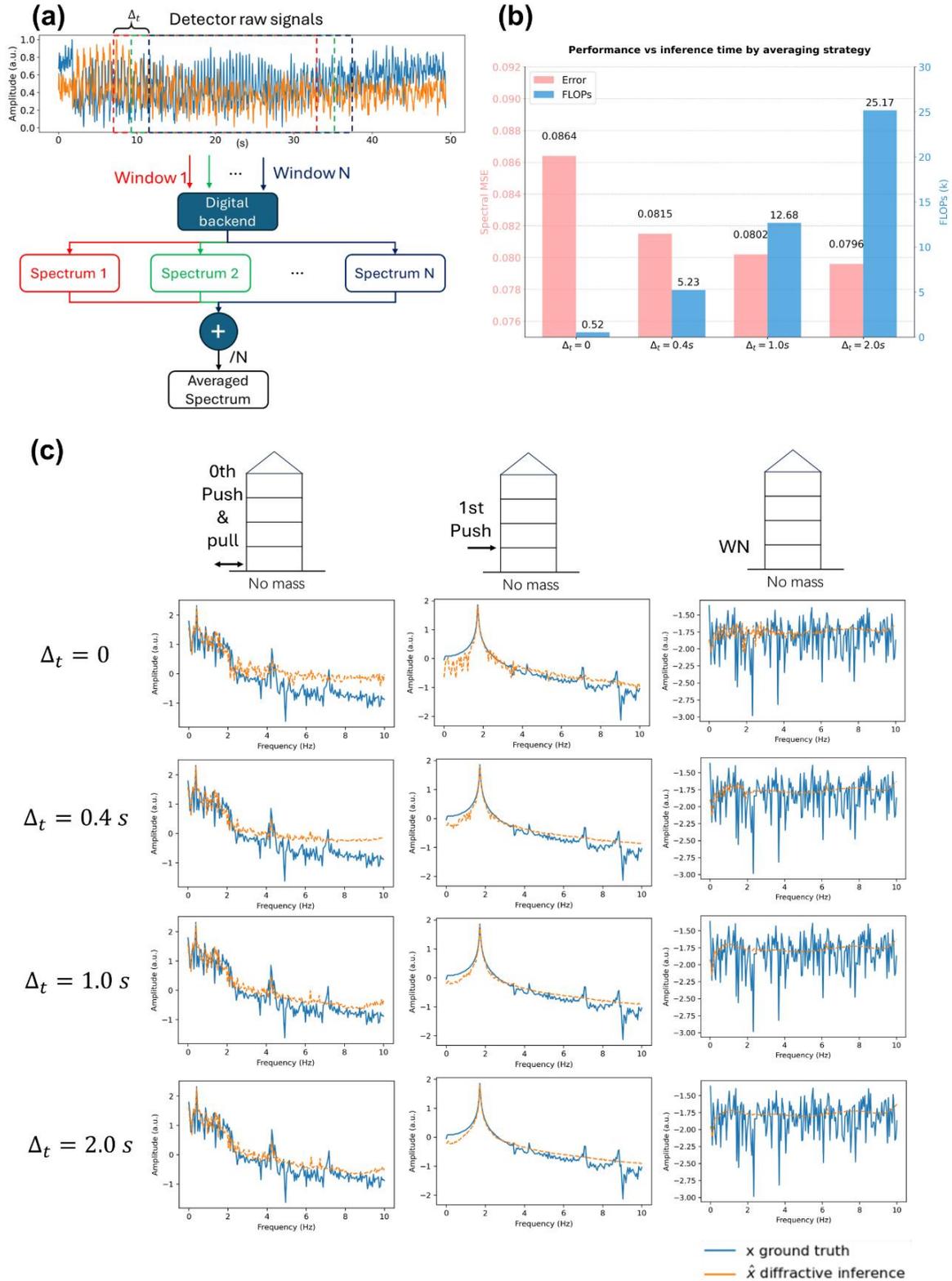

**Figure 7. Temporal averaging improves spectral inference fidelity in diffractive vibration monitoring.** (a) Schematic of the temporal averaging method: raw sensor signals are processed using a sliding window to generate multiple time-series segments. Each segment's spectrum, extracted by the nonlinear digital backend, is then averaged to produce the final spectral output. (b) The trade-off between the spectral inference accuracy and the computational cost. (c) Comparison of the extracted spectra (amplitude vs.

frequency) for different temporal averaging times ($\Delta_t$) under three different experimental conditions. WN: white noise. The non-zero energy at zero frequency was caused by the displacement of the base level ($0^{th}$ level) due to manual perturbation.

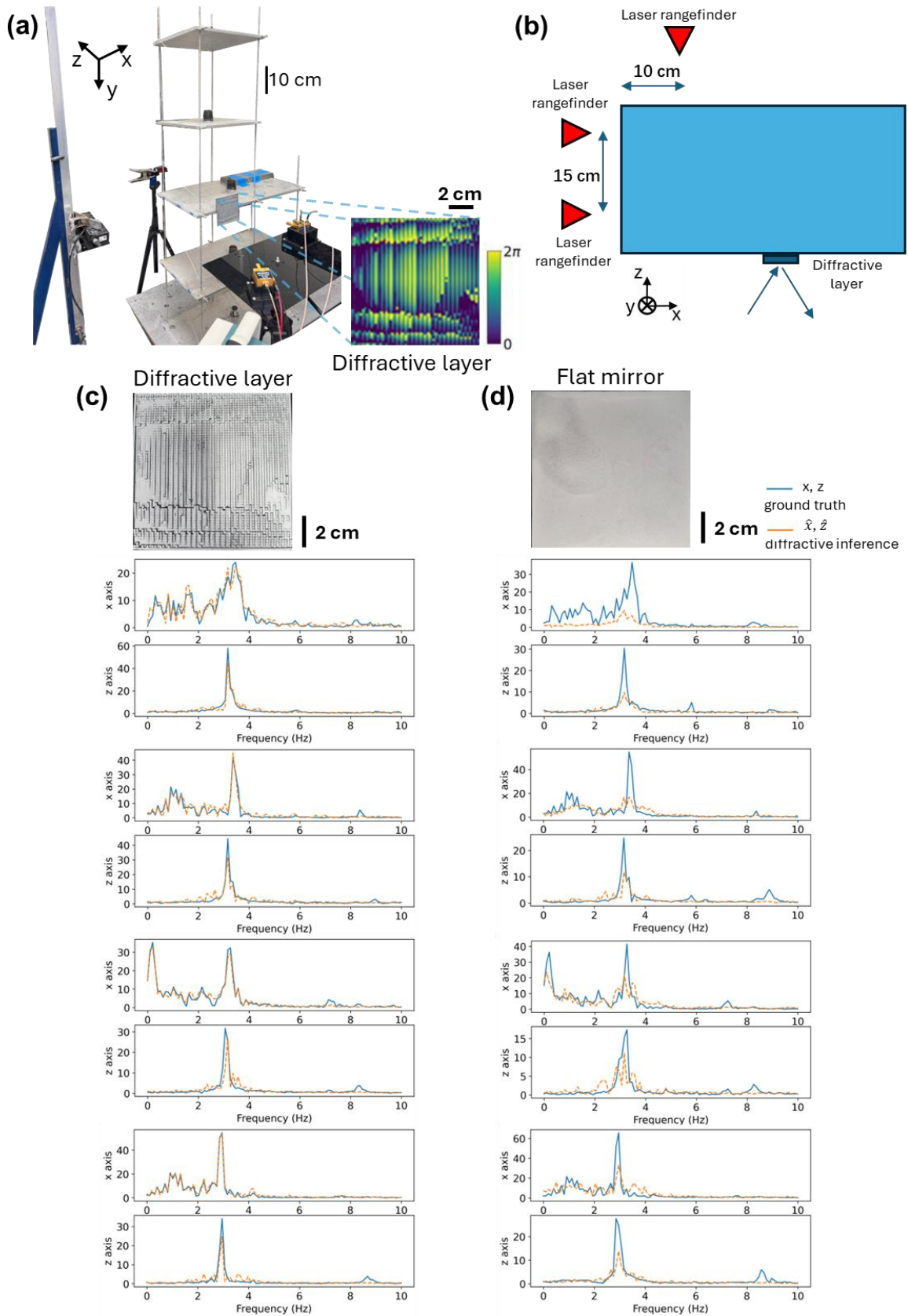

**Figure 8. Experimental validation of the 2D vibration monitoring system using a millimeter-wave source.** (a) Photograph of the experimental setup for 2D vibration spectrum inference. (b) Top-down schematic view showing the placements of the laser rangefinders (used for ground truth measurements) and the diffractive layer. Experimental performance comparison of the 2D diffractive vibration monitoring system between (c) the optimized diffractive layer and (d) a reflective flat mirror that are both 3D printed. The non-zero energy at zero frequency was caused by the displacement of the base level ($0^{th}$ level) due to the manual perturbation.